\documentstyle[12pt]{article}          
\begin{document}
\baselineskip=0.7cm
\renewcommand{\theequation}{\arabic{section}.\arabic{equation}}
\newcommand{\be}{\begin{equation}}
\newcommand{\ee}{\end{equation}}
\newcommand{\bea}{\begin{eqnarray}}
\newcommand{\eea}{\end{eqnarray}}
\newcommand{\nono}{\nonumber}
\newcommand{\meas}[1]{\mbox{d}#1}
\newcommand{\integ}[1]{\int \mbox{d}#1}
\newcommand{\pathint}[1]{\int[\mbox{d}#1]} 
\newcommand{\Tr}{\mbox{Tr}}
\newcommand{\del}[1]{\partial_{#1}}
\newcommand{\vphi}{\varphi}
\newcommand{\sph}{\hat{Y}_{lm}}
\newcommand{\mat}[1]{\mbox{Mat}_{#1}}
\newcommand{\cs}{{\cal C}(S^2)}
\newcommand{\tx}{\tilde{x}}
\newcommand{\ty}{\tilde{y}}
\newcommand{\tu}{\tilde{u}}
\newcommand{\tv}{\tilde{v}}
\newcommand{\argu}{(\theta, \vphi)}
\newcommand{\argup}{(\theta', \vphi')}
\newcommand{\e}{\mbox{e}}
\newcommand{\tlm}{T_{lm}}
\newcommand{\htlm}{\hat{T}_{lm}}
\newcommand{\sint}{\int_{S^2}}
\newcommand{\re}{\mbox{Re}}
\newcommand{\im}{\mbox{Im}}
\newcommand{\ad}{\mbox{ad}}
\newcommand{\Der}{\mbox{Der}}
\newcommand{\D}{\Delta}
\newcommand{\tD}{\tilde{\Delta}}
\newcommand{\tS}{\tilde{S}}
\newcommand{\tf}{\tilde{f}}
\newcommand{\tJ}{\tilde{J}}
\newcommand{\tZ}{\tilde{Z}}
\newcommand{\tW}{\tilde{W}}
\newcommand{\tG}{\tilde{G}}
\newcommand{\te}{\tilde{e}}
\newcommand{\ind}[1]{\mbox{\scriptsize #1}}
\newcommand{\vev}[1]{\langle #1 \rangle}
\newcommand{\bphi}{\bar{\phi}}
\def\thefootnote{\fnsymbol{footnote}}

\begin{flushright}
hep-th/9804041 \\
UT-Komaba/98-9
\end{flushright}

\vspace{1cm}

\begin{center}
\Large 
Master Field on Fuzzy Sphere 

\vspace{1.2cm}
\normalsize
{\sc Tsunehide Kuroki}
\footnote[2]
{\tt kuroki@hep1.c.u-tokyo.ac.jp}

\vspace{0.5cm}
{\it Institute of Physics, University of Tokyo, Komaba, Meguro-ku, 
Tokyo 153, Japan}

\vspace{2.5cm}
\end{center}

\begin{abstract}
The $O(N)$ symmetric vector model is considered on both 
ordinary and fuzzy sphere. It is shown that in both cases 
master fields exist and their explicit forms are presented. 
They are found to mix the internal symmetry and the (fuzzy) space-time 
symmetry. It is also argued that the cutoff brought by the fuzzy sphere 
plays an essential role in constructing the master field.
\end{abstract}
\newpage
\renewcommand{\thefootnote}{\arabic{footnote}}
\section{Introduction}
\setcounter{equation}{0}
\setcounter{footnote}{0}

Noncommutative nature is one of the remarkable features of recent 
developments in nonperturbative string theory. For example, space-time 
coordinates of N D-branes should be treated as non-commuting $N\times N$ 
matrices \cite{boundstate}. In particular, this description of 
coordinates of D-particles exhibits the noncommutative nature of space-time 
at the sub-string scale \cite{shortdistance} where our conventional ideas 
of space-time cease to make sense. Therefore, D-particles are considered 
to be inherently non-local or fuzzy objects. Thus it seems that 
{\em noncommutative geometry} \cite{connes} is a mathematical tool fit 
in the nonperturbative description of string theory.  

The large-$N$ limit is the crucial point other than the noncommutativity 
of space-time in the nonperturbative formulations of string theory. 
It was conjectured \cite{BFSS} that M-theory 
in the infinite momentum frame can be defined as the large-$N$ limit 
of a matrix quantum mechanics obtained from the ten-dimensional SYM theory 
by means of the reduction to $0+1$ dimensional theory. 
It was also proposed \cite{IKKT} that the constructive definition 
of type IIB superstring is given by another matrix model which is the 
large-$N$ reduced model of ten-dimensional SYM theory to a point. 
In relation to the noncommutative structure of space-time, 
it is worth noticing that a similar model to the latter was presented 
\cite{yoneya} as a theory based on the space-time uncertainty principle 
which incorporates a minimal length beyond which the ordinary description 
of space-time by a commutative geometry breaks down. In both cases, 
the large-$N$ limit plays an essential role in matching the degrees of 
freedom of SYM theories with those of M-theory or type IIB theory.\footnote
{Another conjecture is proposed \cite{dlcq} that the sector of DLCQ of 
M-theory is exactly described by a finite-$N$ SYM theory.}

{\em Master field} \cite{witten} is one of the most appealing ideas 
in the context of the large-$N$ field theory.  
It is well-known that the large-$N$ limit of a certain model is governed 
by a single classical field called master field. 
It has the following remarkable features:
\begin{enumerate}
\item The master field links space-time symmetries 
      with the internal symmetry which becomes large in the large-$N$ limit.
\item Physically, the master field may be considered to represent the 
      internal collective motions.  
\end{enumerate} 
Owing to the first feature, space-time dependence of the master field 
is encoded into the internal degrees of freedom 
by making use of the internal symmetry which 
generates the space-time symmetry transformations \cite{gross}. 
In relation to string theory, it has been long pointed out that 
the reduced large-$N$ gauge theories \cite{ek,qek,tek} 
may be solvable and provide a formalism for discussing certain aspects 
of string theory \cite{bars}, and there the master field 
plays a central role \cite{gervaisneveu,migdal}.  
The reduced model proposed in \cite{IKKT} 
is based on the large-$N$ reduction of SYM theory to a point, which is 
nothing other than the master field of the theory. As for the second feature, 
we recall that higher dimensional D-branes may be regarded 
as bound states of D0-branes \cite{town} and collective motions 
of lower dimensional D-branes give rise to fluctuation of higher 
dimensional one. Thus master fields of SYM theory may be relevant 
to dynamics of D-branes.  
     
These observations tempt us to formulate the large-$N$ SYM theory 
on more generic noncommutative geometry and to examine its master field.   
Such a formulation is expected to provide new insights into the 
nonperturbative definition of string theory as in \cite{BFSS,IKKT,yoneya}.
As a first step in this project, it is instructive to formulate a simple 
large-$N$ field theory on the known noncommutative geometry and to 
examine the existence and the property of the master field. 
In particular, it seems interesting to clarify the relation 
mentioned above between the space-time and the internal symmetry 
because in this case the space-time structure is fuzzy. 
In this paper, we take the fuzzy sphere as a noncommutative geometry which 
is defined in \cite{fs} and consider the $O(N)$ symmetric vector model 
on it. 

Another motivation is from the field theoretic point of view. 
The fuzzy sphere can also be regarded as one of the regularization 
schemes which manifestly preserves the rotational symmetry unlike the lattice 
regularization. Moreover, the supersymmetric version of the 
fuzzy sphere (`fuzzy supersphere') is constructed recently \cite{superfs}. 
It should be noticed in that it provides regularization which manifestly 
preserves the supersymmetry which is difficult to preserve in the 
lattice regularization. 
Moreover, also from string theory point of view, if we intend to define 
nonperturbative superstring theory as the large-$N$ limit of a certain model 
as in \cite{BFSS,IKKT}, we expect this model to be manifestly supersymmetric 
and to be regularized by a minimal length because string theory is 
ultraviolet finite due to a minimal length. Noncommutative geometry 
including the fuzzy sphere naturally incorporates such a minimal length. 
Thus noncommutative geometry could be again the geometrical framework 
in which nonperturbative superstring theory should be described. 
However, only a few examples 
have been explored for field theories on the noncommutative geometry. 
Thus we believe that our model is useful to deepen our 
understanding of the field theory on the noncommutative geometry and 
its role as regularization of the theory.

The organization of this paper is as follows: in the subsequent section, 
we give a brief review of the formulation based on the idea of the master 
field. In section 3 we review the definition of the fuzzy sphere and 
of a field theory defined on it. In section 4, we consider the $O(N)$ 
symmetric vector model on the ordinary sphere as a commutative warm-up 
for examining the same model on the fuzzy sphere. 
We will give the explicit representation for the master field. 
Section 5 is a main part of this paper. There the $O(N)$ symmetric 
vector model is considered on the fuzzy sphere. It is shown that 
this model also has the `master field'. It is also seen 
that regularization by means of the fuzzy sphere is essential for 
the construction of the master field. The last section contains the 
conclusions and the discussions of further extensions of our work.   
\section{Master Field in the Large-$N$ Field Theory}
\setcounter{equation}{0}

In this section, we consider the generic large-$N$ field theory 
and present a formulation of it based on the idea of the master field  
following \cite{itoyone}. 

We begin by the Euclidean path integral representation 
for the partition function
\begin{equation}
Z=\pathint{\phi}\exp \left( -\frac{1}{g^2}S[\phi]\right). 
\end{equation}
This can be formally written as 
\begin{equation}
Z=\integ{S}\exp \left[ -\frac{1}{g^2}(S-g^2{\cal J}(S))\right], \label{Z}
\end{equation}
\begin{equation}
{\cal J}(S)=\ln \pathint{\phi} ~\delta (S-S[\phi]).
\end{equation}
The entropy factor ${\cal J}(S)$ measures the volume of the action orbit
${\cal O}(S)$ which is a set of configurations with a given value (=$S$) 
of the action functional. 
Suppose that the action is invariant under some internal symmetry 
transformation whose degrees of freedom increase as $N$ becomes large. 
By definition, for any configurations $\bar{\phi}$, 
the symmetry orbit ${\cal O}_I(\bar{\phi})$ which consists of configurations 
given by the internal symmetry transformations of $\bar{\phi}$ 
must be included in ${\cal O}(S[\bar{\phi}])$.
If we take the large-$N$ limit, the volume of the symmetry orbit 
increases in general and it is possible that the symmetry orbit effectively 
covers the whole action orbit for a suitable choice of $\bar{\phi}$. 
Then the entropy can be computed only from the transformation property 
of the field:
\be
{\cal J}(S[\bar{\phi}])\sim \ln\mbox{vol}({\cal O}(S[\bar{\phi}]))
                       \sim \ln\mbox{vol}({\cal O}_I(\bar{\phi})).
\ee
We call such configurations {\em maximal entropy configurations}.   
In many cases, one can take the large-$N$ limit so that 
$S_{\mbox{\scriptsize{eff}}}=S-g^2{\cal J}(S)$ increases 
in proportion to a positive power of $N$.  There $\bar{\phi}$ can be given 
as a solution to the saddle-point equation. 
This suggests the existence of a single dominant symmetry orbit 
for describing all amplitudes in the large-$N$ limit which is 
a candidate for the master field. 

Based on the above idea, one can derive the equation of the master field 
$\bar{\phi}$ \cite{itoyone}, 
\bea 
\frac{\delta S[\bar{\phi}]}{\delta \bar{\phi}_i(x)}
+g^2\sum_{\alpha,\beta}M^{-1\alpha \beta}[\bar{\phi}]
(\delta^{\alpha}\delta^{\beta}\bar{\phi}(x))_i=0, \label{mastereq} \\
M^{\alpha\beta}[\bphi]
=\int (\delta^{\alpha}\bphi(x))_i(\delta^{\beta}\bphi(x))_i\meas{x},
\eea
where $\delta^{\alpha}$ denotes the infinitesimal internal symmetry 
transformation.
\section{The Fuzzy Sphere}
\setcounter{equation}{0}
\setcounter{footnote}{0}

 In this section, we briefly discuss the main characteristics of the fuzzy 
sphere and an example of a field theory on it. In what follows, 
we distinguish quantities defined on the ordinary sphere 
from those defined on the fuzzy sphere by denoting the former 
with the tilde. 
These arguments are largely based on \cite{fs,hoppe}.  

\subsection{Definitions}

In this subsection we consider the sequence of algebras $\mat{M}$
of $M\times M$ complex matrices and find that in the large-$M$ limit
these geometries tend towards the algebra ${\cal C}(S^2)$ of smooth 
complex-valued functions on $S^2$. In other words, finite dimensional 
algebra $\mat{M}$ for each $M$ can be regarded as finite truncation 
or regularization of infinite dimensional algebra ${\cal C}(S^2)$. 
We also see that this regularization corresponds to an ultraviolet cutoff.
Since $\mat{M}$ is of course a noncommutative algebra,
the base space can be considered to be noncommutative.
We refer such a space to {\em the fuzzy sphere}.

Consider ${\bf R}^3$ with coordinates $\tx^a (a=1,2,3)$ and the 
standard Euclidean metric $g_{ab}=\delta_{ab}$. 
The ordinary sphere is defined by 
\be
\delta_{ab}\tx^a\tx^b=r^2.\label{sphere}
\ee
We associate these coordinate functions $\tx^a$ with the generators of 
$SU(2)$: 
\be
\tx^a \mapsto x^a \equiv \kappa J^a,
\ee 
where $J^a$ be an $M$-dimensional irreducible representation\footnote
{We need this irreducibility to make the derivations
of $\mat{M}$ complete. See below.}
(spin $J=(M-1)/2$ representation) of the Lie algebra of 
$SU(2):~~[J^a, J^b]=i\epsilon_{abc}J^c$, and $\kappa$ be
a positive number defined by $4r^2=(M^2-1)\kappa^2$ from which it follows 
that 
\be
\delta_{ab}x^a x^b =r^2\cdot\mbox{\bf 1}.
\ee 
Let ${\cal P}$ be the algebra of analytic functions of $\tx^a$ 
and ${\cal I}$ be its ideal consisting of all functions of a form 
$h(\tx^a)(\sum_a \tx^{a^2}-r^2)$. Then the quotient algebra 
${\cal P}/{\cal I}$ is dense in the algebra $\cs$. 
Any element $\tilde{f} \in {\cal P}/{\cal I}$ can be represented as 
\be
\tilde{f}=\sum_{l=0}^{\infty}\frac{1}{l!}f_{a_1\cdots a_l}
          \tx^{a_1}\cdots\tx^{a_l}, 
\label{funexp}
\ee
where $f_{a_1\cdots a_l}$ is traceless between any two indices and totally
symmetric. Truncating the expansion at terms of order $M-1$ 
and replacing $\tx^a$ with $x^a$, we get an element of $\mat{M}$ 
corresponding to $\tf$, 
\be
f=\sum_{l=0}^{M-1}\frac{1}{l!}f_{a_1\cdots a_l}
                        x^{a_1}\cdots x^{a_l} \in \mat{M}.
\label{matexp}
\ee
This is the truncation map $\phi_M:\cs\rightarrow\mat{M}$. 
If we define a quantity $k\equiv 2\pi\kappa r$ which has the dimension of 
$(\mbox{length})^2$, it plays a role of a minimal length
in the sense that the generators $x^a$ of the algebra $\mat{M}$ satisfy
\be
[x^a, x^b]=i\frac{k}{2\pi}C_{abc}x^c,~~~~~~~~~
C_{abc}=\frac{1}{r}\epsilon_{abc}.
\ee
This means that for a finite $M$ the `coordinates' $x^a$ of the fuzzy 
sphere are noncommutative, while in the limit $M \rightarrow \infty$, 
they commute with each other and all of the points of $S^2$ can be 
distinguished. So the ordinary $S^2$ can be recovered.
 
Next let us see that this truncation provides an ultraviolet cutoff 
for theories on the fuzzy sphere. If we define a norm on $\cs$ as 
\be
||\tilde{f}||^2\equiv \frac{1}{4\pi r^2}\sint |\tilde{f}|^2,
\label{scalarproductoncs2}
\ee
where $\sint$ is an abbreviation for 
$\int r^2\sin\theta\meas{\theta}\meas{\vphi}$, then one can take 
as an orthonormal basis the usual spherical harmonics 
$\{\sph(\theta,\vphi)\}$ ($l\geq 0, -l\leq m \leq l$) defined by
\be
\sph\argu \equiv \sqrt{4\pi}Y_{lm} \equiv 
\sqrt{4\pi}N_{lm}P_{l|m|}(\cos\theta)~\e^{im\vphi},
\label{rescaledylm}
\ee
where $Y_{lm}\argu$ is the standard spherical harmonics,  $N_{lm}$ is 
its normalization constant,
\be
N_{lm}\equiv (-)^{\frac{m+|m|}{2}}\sqrt{\frac{(2l+1)}{4\pi}
\frac{(l-|m|)!}{(l+|m|)!}}
\ee
and $P_{lm}(\cos\theta)$ is an associated Legendre function.
Here we note that solid harmonics $r^l\sph\argu$ is homogeneous polynomial of
degree $l$ in variables $\tx^1=r\sin\theta\cos\vphi, 
\tx^2=r\sin\theta\sin\vphi$, and $\tx^3=r\cos\vphi$.
Therefore, $\{\sph \}$ ($0\leq l \leq M-1, -l\leq m\leq l$) form a basis 
of the vector space which consists of functions on $S^2$ which can be 
expanded in terms of $\tx^a$ of order up to and including $M-1$. 
Thus we have observed that the truncation to $\mat{M}$ is nothing but 
a cutoff of higher angular momenta $l\geq M$.

For later convenience, we define an orthonormal basis of $\mat{M}$ 
corresponding to $\sph$. A norm on $\mat{M}$ corresponding to 
(\ref{scalarproductoncs2}) is defined as 
\be
||f||_M^2\equiv \frac{1}{M}\Tr(f^{\dagger}f).
\ee
Then there exists an orthonormal basis $\{\tlm\}$ ($0\leq l\leq M-1, 
-l\leq m \leq l$) 
\be
||\tlm T_{l'm'}||_M=\frac{1}{M}\Tr(\tlm^{\dagger}T_{l'm'})
                   =\delta_{ll'}\delta_{mm'},
\ee
which tends to $\sph$ in the large-$M$ limit. 

So far we have discussed that $\mat{M}$ determines noncommutative
geometry (the fuzzy sphere) and provides regularization of the ultraviolet 
divergences like a lattice field theory. 
However, unlike lattice regularization, 
it manifestly preserves the rotational symmetry $SO(3)$ of space-time, 
because $\tlm$ for fixed $l$ and $-l\leq m \leq l$ form 
a $(2l+1)$-dimensional representation of $SO(3)$. 
Namely, for an $SO(3)$ rotation $R$, 
\be
U(R)\tlm U(R)^{-1}=\sum_{m'=-l}^l T_{lm'} R_{mm'}^l(R),
\label{tlmrotation}
\ee
where $U(R)$ is a $M$-dimensional representation of $R$ and
$R_{mm'}^l(R)$ are the rotation matrices for angular momentum $l$.

On $S^2$ we have vector fields 
\be
\te_a=-C_{abc}\tx^b\tilde{\partial}_c,
\ee
which satisfy 
\be
\tD=\delta^{ab}\te_a\te_b,
\label{spherelaplacian}
\ee
where $\tD$ is the ordinary Laplacian on $S^2$ of radius $r$. 
It is argued in \cite{fs} that we can correspondingly define 
vector fields on the fuzzy sphere, i.e., derivations of $\mat{M}$
as
\be
e_a=\frac{2\pi}{ik}\ad(x^a),~~~~~~~~~ 
\label{derivation}
\ee
and Laplacian on the fuzzy sphere as 
\be
\D=\delta^{ab}e_ae_b,
\label{matlaplacian}
\ee 
which are shown to become $\te_a$ and $\tD$ in the large-$M$ limit 
respectively. The set of three derivations $e_a$ are complete in the sense 
that if $e_af$=0, then $f$ must be proportional to the unit matrix, 
which follows from the irreducibility of the representation of $SU(2)$
which defines $\mat{M}$. 
As $\sph$ ($-l\leq m \leq l$) are eigenfunctions of $\tD$: 
\be
\tD\sph=\frac{l(l+1)}{r^2}\sph,
\ee 
so $\tlm$ ($-l\leq m \leq l$) are eigenmatrices of $\D$: 
\be
\D\tlm=\frac{l(l+1)}{r^2}\tlm.
\ee

\subsection{Field Theory}

The ordinary real scalar field theory on the Euclidean sphere is defined 
by the partition function
\be
\tilde{Z}[\tilde{J}]=\int_{\cs}[\meas{\tilde{f}}]~
\e^{-\tilde{S}(\tilde{f},\tilde{J})},
\label{partitionfun}
\ee
where $\tilde{J}$ is an appropriate external source and 
$\tilde{S}$ is an action
\be
\tS(\tf,\tJ)=\sint \left(-\frac{1}{2}\tf\tD\tf+\frac{1}{2}\mu_0^2\tf^2
            +V_{\ind{int}}(\tf)+\tJ\tf\right).
\label{usualaction}
\ee  
The functional integral $\int_{\cs}[\meas{\tilde{f}}]$ is performed 
over the infinite-dimensional space $\cs$ and requires some regularization 
to make it well-defined.     
As discussed in the previous subsection, the fuzzy sphere is 
considered as a way of the ultraviolet regularization. Namely, 
replacing $\cs$ with $\mat{M}$ by the truncation map $\phi_M$, 
we consider corresponding to (\ref{partitionfun}),
\be 
Z_M[J]=\int_{\ind{$\mat{M}$}}\meas{f}~\e^{-S(f,J)},
\ee
\be
S(f,J)=k\Tr\left(-\frac{1}{2}f\D f+\frac{1}{2}\mu_0^2f^2+V_{\ind{int}}(f)
      +Jf\right),
\label{fuzzyaction}
\ee
where $f,J$ are Hermitian $M\times M$ matrices
which are the image of $\tilde{f},\tilde{J}$ by $\phi_M$ respectively, 
and $\int_{\ind{$\mat{M}$}}\meas{f}$ is understood to denote the integration 
over all of the components of the matrix $f$. Then for each $M$,  
$Z_M[J]$ is well-defined and is expected to become (\ref{partitionfun}) 
in the large-$M$ limit.
Moreover, $Z_M$ is in itself regarded as the definition of the quantum 
field theory on the fuzzy sphere. 
  
Next let us give the definition of the correlation 
functions of the theory (\ref{fuzzyaction}). 
On the analogy of the commutative case, we adapt the following definitions 
for the correlation functions and the propagator:
\bea
\vev{f_1f_2\cdots f_n}_J & \equiv & Z[J]^{-1}
\frac{\delta^n Z[J]}{\underbrace{k\delta J\cdots k\delta J}_{n~\ind{times}}}~
\in                            
\underbrace{\mat{M}\otimes\cdots\otimes\mat{M}}_{n~\ind{times}}, \\
G & \equiv & \frac{\delta^2 \ln W[J]}{k\delta Jk\delta J}~
\in \mat{M}\otimes\mat{M}.
\eea
Note that differentiating $n$ times the function of $\mat{M}$ 
with respect to its element yields an element of the direct product 
of $n$ copies of $\mat{M}$. For example, if $V_{\ind{int}}=0$, 
the free propagator (Green function) is given as 
\be
G_0=\frac{1}{kM}\sum_{l=0}^{M-1}\sum_{m=-l}^l\frac{\tlm\otimes\tlm^{\dagger}}
                                               {\frac{l(l+1)}{r^2}+\mu_0^2},
\label{fuzzyfreepropagator}
\ee
which tends to the Green function on the sphere 
\be
\tilde{G}_0(x,y)=\frac{1}{4\pi r^2}\sum_{l=0}^{\infty}\sum_{m=-l}^{l}
                 \frac{\sph (x)\sph^{\ast}(y)}
                      {\frac{l(l+1)}{r^2}+\mu_0^2},
\label{freepropagator}
\ee
in the large-$M$ limit.

Let us conclude this section by a comparison of two theories regularized 
by a naive angular momentum cutoff and by the fuzzy sphere. 
Although both theories preserve the rotational symmetry manifestly, 
it makes an important difference that the former apparently fails to 
preserve the angular momentum while the `fuzzy angular momentum' 
conservation law holds in the latter in the sense that $\{\tlm\}$ 
($0\leq l\leq M-1$, $-l\leq m\leq l$) form a closed algebra.  
\section{$O(N)$ Symmetric Vector Model on the Sphere}
\setcounter{equation}{0}
\setcounter{footnote}{0}

In this section, the $O(N)$ symmetric vector model is considered on the 
sphere of the radius $r$. Using the results in section 2, 
we construct the large-$N$ 
master field explicitly and show that it in fact reproduces the $O(N)$
invariant two-point function. Of course, these results can be regarded 
as the commutative limit of those of the $O(N)$ symmetric vector model 
on the fuzzy sphere considered in the next section.

\subsection{Large-$N$ Limit}

The action we consider is
\be 
S=\int_{S^2} \left(-\frac{1}{2}\phi_i\tilde{\Delta}\phi_i
 +\frac{1}{2}\mu_0^2 \phi_i^2+\frac{g_0}{4N}(\phi_i^2)^2\right),
\label{spherevectormodel}
\ee
where $\tilde{\Delta}$ is the Laplacian on the sphere given
in (\ref{spherelaplacian}) and $\phi_i$ is a real scalar field 
which transforms as a component of a $N$-dimensional vector 
under the $O(N)$ rotation.
Note that the sphere is imbedded in ${\bf R}^3$ with the Euclidean metric 
so that it is possible to apply the formulation 
introduced in section 2 and to extend it to the fuzzy sphere.\footnote
{There is at present no satisfactory noncommutative version
of Minkowski space.} 

Introducing the auxiliary field $\sigma\argu$ and performing the Gaussian 
integration over the $N$-component field $\phi_i$, we obtain the following 
expression for the partition function,
\be
Z=\int {\cal D}\sigma\e^{-\frac{N}{2}S_{\mbox{\scriptsize{eff}}}}, 
\ee
\be
S_{\mbox{\scriptsize{eff}}}=\sint\left(-\frac{1}{2}g_0\sigma^2
                           +\frac{1}{4\pi r^2}\log
                            \frac{\det(-\tilde{\Delta}+\mu_0^2+g_0\sigma)}  
                                 {\det(-\tilde{\Delta}+\mu_0^2)}\right),
\ee
where we added a constant to $S_{\ind{eff}}$ so that 
$S_{\ind{eff}}|_{\sigma=0}=0$. In the large-$N$ limit, 
the dominant contribution to the integral comes from 
the rotationally invariant saddle point 
$\sigma\argu=\sigma$ which is a solution of the saddle point equation
\be
0=\frac{\partial S_{\mbox{\scriptsize{eff}}}}{\partial\sigma}
 =-g_0\sigma+\frac{g_0}{4\pi r^2}
  \Tr\frac{1}{-\tilde{\Delta}+\mu_0^2+g_0\sigma},
\label{saddlepoint}
\ee
and hence the gap equation
\be
\sigma=\frac{1}{4\pi r^2}\sum_{l=0}^L\sum_{m=-l}^l
       \frac{1}{\frac{l(l+1)}{r^2}+\mu_0^2+g_0\sigma}\nono \\
      =\frac{1}{4\pi r^2}\sum_{l=0}^L
       \frac{2l+1}{\frac{l(l+1)}{r^2}+\mu_0^2+g_0\sigma},
\label{spheregapeq}
\ee
where we have introduced the cutoff $L \in {\bf Z}$ for the angular momentum.

Next let us calculate the two-point function (propagator)
$\tilde{G}_{ij}(x,y;\mu_0^2)=\langle \phi_i(x)\phi_j(y) \rangle$ 
($x,y \in S^2$) in the large-$N$ limit.
Making the Feynman graph expansion shows that the leading contributions 
to the propagator are ``tree chain diagrams'', namely, randomly 
branching polymers. Thus let $\sigma_0$ be the sum of all connected
tree chain diagrams, then we can express $\tilde{G}_{ij}(x,y;\mu_0^2)$
as
\be
\tilde{G}_{ij}(x,y;\mu_0^2)
     =\frac{\delta_{ij}}{4\pi r^2}\sum_{l=0}^L\sum_{m=-l}^l
      \frac{\sph(x)\sph^{\ast}(y)}{\frac{l(l+1)}{r^2}+\mu_0^2+g_0\sigma_0}.
\label{spheretwopointfun}  
\ee
Consistency condition on $\sigma_0$ requires that
$\sigma_0$ is given as a solution to the gap equation.

\subsection{ Master Field on the Sphere} 

Now let us apply the formulation developed in section 2 to our model
and derive the master field.
In the present case, the master field equation of motion 
eq.(\ref{mastereq}) becomes 
\bea
0 & = & (-\tD+2V'(\phi(x)^2))\phi_i(x)
    +\sum_{\alpha,\beta}M^{-1\alpha\beta}[\phi]
     (\tau^{\alpha}\tau^{\beta}\phi(x))_i, \label{spheremastereq} \\
  & & M^{\alpha\beta}=-\sint\phi_i(x)(\tau^{\alpha}\tau^{\beta}\phi(x))_i,~~~~
      \tau^{\alpha}:\mbox{infinitesimal generator of $O(N)$} \nono
\eea
where 
\be
V(x)=\frac{1}{2}\mu_0^2 x+\frac{g_0}{4N}x^2.
\label{potentialform}
\ee
In order to solve this equation, we adopt a following ansatz: \\ 
in an appropriate base, the master field (i.e. a solution to the 
above equation) $\bphi_i(x)$ is given by  

for $l^2 < i < (l+1)^2~(l\in {\bf N}\cup\{0\})$,
\be
\bphi_i\argu=\left\{\begin{array}{ll}
         \sqrt{2}c_l\re \hat{Y}_{l~\frac{1}{2}\left\{(l+1)^2-i\right\}}\argu 
        & \mbox{if $i-l^2:~$odd} \\          
         \sqrt{2}c_l\im \hat{Y}_{l~\frac{1}{2}\left\{(l+1)^2-(i-1)\right\}}
         \argu 
        & \mbox{if $i-l^2:~$even},          
                   \end{array}
            \right.
\label{spheremasterdef1}
\ee

and for $i=(l+1)^2$,
\be
\bphi_{(l+1)^2}\argu=c_l \hat{Y}_{l0}\argu,
\label{spheremasterdef2}
\ee
where $\sph\argu$ is spherical harmonics defined in section 3.
Note that $\hat{Y}_{l0}=\hat{Y}_{l0}^{\ast}\in {\bf R}$.
To be more explicit, 
\be
\bphi=\sqrt{2}\left(\begin{array}{c}
                    c_0/\sqrt{2}\hat{Y}_{00} \\ \hline 
                    c_1\re\hat{Y}_{11} \\
                    c_1\im\hat{Y}_{11} \\
                    c_1/\sqrt{2}\hat{Y}_{10} \\ \hline 
                    c_2\re\hat{Y}_{22} \\
                    c_2\im\hat{Y}_{22} \\
                    c_2\re\hat{Y}_{21} \\
                    c_2\im\hat{Y}_{21} \\
                    c_2/\sqrt{2}\hat{Y}_{20} \\ \hline
                       .          \\
                       .          \\
                       .          
                   \end{array}
             \right).
\label{spheremaster}
\ee
Note that $\bphi_i\argu$ is a rotational invariant field in the sense that 
there exists an orthogonal matrix $O(\theta', \vphi') \in O(N)$ 
such that 
\be
\bphi_i(\theta+\theta', \vphi+\vphi')=O_{ij}(\theta', \vphi')\bphi_j\argu.
\label{claim}
\ee

 
Substituting the ansatz into the master field equation of motion 
(\ref{spheremastereq}) and picking up the $l^2 < i\leq (l+1)^2$ 
($l\in {\bf N}\cup \{0\}$) component yields
\be
0=\left(\frac{l(l+1)}{r^2}+2V'(\bphi^2)\right)\bphi_i
 +\sum_{\alpha\beta}M^{-1\alpha\beta}[\bphi](\tau^{\alpha}\tau^{\beta}\bphi)_i.
\label{substituteresult}
\ee
Here it is worth noticing that $\bphi_i^2\argu$ is in fact 
independent of $\theta, \vphi$: 
\be
\bphi_i^2\argu=\sum_{l}c_l^2(2l+1),
\label{point}
\ee
where the well-known formula for the spherical harmonics 
$\sph\argu$ is employed:
\be
\sum_{m=-l}^l\sph^{\ast}\argu\sph\argu=2l+1.
\label{wellknownformula}
\ee
This relation enables us to rewrite the master field equation of motion 
as
\be
\left(\frac{l(l+1)}{r^2}+2V'(\sum_l (2l+1)c_l^2)\right)\bphi_i
+\sum_{\alpha\beta}M^{-1\alpha\beta}[\bphi]
(\tau^{\alpha}\tau^{\beta}\bphi)_i=0.
\label{rewrittenmastereq}
\ee

To make the field given by the ansatz a maximum-entropy configuration, 
the set of $(l,m)$ must span the whole angular-momentum number and 
magnetic quantum number space without degeneracy in the large-$N$ limit. 
Then we have 
\be
M^{\alpha\beta}=-\sint\bphi_i\argu(\tau^{\alpha}\tau^{\beta}\bphi)_i\argu
               =-4\pi r^2(c_l^2+c_k^2)\delta^{\alpha\beta},
\label{projection}
\ee
where the index $\alpha$ is understood to denote the $O(N)$ rotation 
in the $ij$-plane, and $l^2<i\leq (l+1)^2, k^2<j\leq(k+1)^2$. 
Thus picking up the component for $l^2 < i\leq(l+1)^2$, 
eq.(\ref{rewrittenmastereq}) takes the form 
\be
\left(\frac{l(l+1)}{r^2}+2V'(\sum_k (2k+1)c_k^2)\right)c_l
=\frac{1}{4\pi r^2}\left(\sum_k\frac{(2k+1)c_l}{c_l^2+c_k^2}
                              -\frac{1}{2c_l}\right).
\ee
We find in the large-$N$ limit, $c_l$ is given by 
\be
c_l^2=\frac{N}{4\pi r^2}
      \frac{1}{\frac{l(l+1)}{r^2}+2V'(\sum_l(2l+1)c_l^2)}.
\label{coefficient}
\ee
Thus we obtain the gap equation in the following form: 
\bea
\sigma & = & \frac{1}{4\pi r^2}
             \sum_{i=1}^{N}\frac{1}{\frac{l(l+1)}{r^2}+2V'(N\sigma)} \nono \\
       & = & \frac{1}{4\pi r^2}\sum_l
             \frac{2l+1}{\frac{l(l+1)}{r^2}+\mu_0^2+g_o\sigma}.
\eea
where $\sigma=\sum_l(2l+1)c_l^2/N$. 
This indeed coincides with the gap equation (\ref{spheregapeq}) 
derived by means of the saddle point method. 
Using (\ref{spheregapeq}), (\ref{spheretwopointfun}), 
and (\ref{coefficient}), we can also verify that the field given 
by (\ref{spheremaster}) reproduces the $O(N)$ invariant two-point function 
\be
\frac{1}{N}\sum_i\vev{\phi_i(x)\phi_i(y)}
=\frac{1}{4\pi r^2}\sum_{l=0}^L\sum_{m=-l}^l
 \frac{\sph(x)\sph^{\ast}(y)}{\frac{l(l+1)}{r^2}+\mu_0^2+g_0\sigma_0}
=\frac{1}{N}\sum_i\bphi_i(x)\bphi_i(y).  
\ee
Thus (\ref{spheremaster}) explicitly realizes the master field of 
the $O(N)$ symmetric vector model on the sphere. To the best of our 
knowledge, it is the first example of the master field of linear 
$\sigma$-model on the sphere. 

We mention some remarks concerning the master field (\ref{spheremaster}). 
As we mentioned in the introduction, it mixes the internal and space-time 
rotational symmetry. In fact, we have one-to-one correspondence 
between the internal symmetry index $i$ and the pair 
of the angular momentum and the magnetic quantum number $(l,m)$ 
within the master field (\ref{spheremasterdef1}), (\ref{spheremasterdef2}). 
It can be regarded as one of the consequence of (\ref{claim}). 
Next we note the relationship between the existence 
of the master field on the sphere and the cutoff of the theory.
In view of the form of (\ref{spheremaster}), 
it is clear that the angular momentum cutoff $L$ is closely related to $N$.    
Therefore, we encounter a subtlety in taking 
the large-$N$ limit and the limit $L\rightarrow\infty$. 
In the next section, we show that considering the theory 
on the fuzzy sphere settles this problem.
\section{Master Field on the Fuzzy Sphere} 
\setcounter{equation}{0}
\setcounter{footnote}{0}

In this section, we consider the master field for the $O(N)$ 
symmetric vector model on the fuzzy sphere and discuss its large-$N$ 
limit. Then following the formulation in section 2, it is shown that 
this model also has a master field similar to (\ref{spheremaster}). 

\subsection{$O(N)$ Symmetric Vector Model on the Fuzzy Sphere} 

Combining the field theory (\ref{fuzzyaction}) presented 
in section 3 and the vector model (\ref{spherevectormodel}) in section 4, 
we consider the $O(N)$ symmetric vector model on the fuzzy sphere 
defined by the action
\be
S=k\Tr\left[-\frac{1}{2}\phi_i\D\phi_i
 +\frac{1}{2}\mu_0^2\phi_i^2+\frac{g_0}{4N}(\phi_i^2)^2\right],
\ee
where $\phi_i$ is an $M\times M$ Hermitian matrix for each $i$ and 
transforms under the $O(N)$ rotation as 
\be
\phi_i\rightarrow\phi'_i=O_{ij}\phi_j,~~~O\in O(N).
\ee
This action has an above $O(N)$ rotational symmetry as well as 
an $SU(2)$ symmetry 
\be
\phi_i\rightarrow\phi'_i=\phi_i+\varepsilon e_a\phi_i
                        =\phi_i+\frac{2\pi\varepsilon}{ik}[x_a,\phi_i],
\ee   
which is a remnant of the conformal symmetry on the sphere. 

The equation of motion is 
\be
(\D-\mu_0^2)\phi_i=\frac{g_0}{2N}(\phi_i(\phi_j^2)+(\phi_j^2)\phi_i), 
\ee
and the free propagator is $G_{0ij}=\delta_{ij}G_0$ 
where $G_0$ is given in (\ref{fuzzyfreepropagator}).

Let us calculate the two-point function in the large-$N$ limit 
according to the definition in section 3. For this purpose, 
introducing an external source of Hermitian matrix $J_i\in\mat{M}$ 
for each $i$ and an auxiliary Hermitian matrix $\sigma\in\mat{M}$ 
in the same way as in sections 3, 4, the action becomes 
\be
S(\phi_i,\sigma,J_i)=k\Tr\left[-\frac{1}{2}\phi_i\D\phi_i
                               +\frac{1}{2}\mu_0^2\phi_i^2
                               +J_i\phi_i-\frac{1}{4}Ng_0\sigma^2
                               +\frac{1}{2}g_0\phi_i^2\sigma
                         \right].
\ee
The classical field $\phi_i^c$ is a solution to the equation 
of motion derived from this action
\be
(\D-\mu_0^2)\phi_i^c=J_i+\frac{1}{2}g_0(\phi_i^c\sigma+\sigma\phi_i^c).
\label{fuzzyvectoreom}
\ee
Expanding the action around $\phi_i^c$ and integrating over fluctuations, 
we obtain the partition function in the presence of the source 
\be
Z[J]=(2\pi)^{\frac{NM^2}{2}}\int\meas{\sigma}
     \left(\det (-k(-\D+\mu_0^2+g_0\sigma))\right)^{-\frac{N}{2}}
     \exp\left[-k\Tr\left(\frac{1}{2}J_i\phi_i^c-\frac{1}{4}Ng_0\sigma^2
                    \right)
         \right]. 
\label{fuzzyvectorpartitionfun}
\ee
In the large-$N$ limit, $\sigma$ in (\ref{fuzzyvectorpartitionfun}) 
is reduced to a matrix proportional to the unity which is determined 
as a `rotationally invariant' saddle point of an effective action 
(see below) in the sense that $e_a\sigma=0$. Note that as mentioned 
in section 3, the irreducibility of the representation of $x^a$ is 
important for the existence of such a matrix.  
Thus given such a  scalar matrix $\sigma_0$, $\phi_i^c$ is the solution 
to the equation of motion in the large-$N$ limit, 
\be
(\D-\mu_0^2)\phi_i^c=J_i+g_0\sigma_0\phi_i^c,
\ee
being given by 
\be
\phi_i^c=-k\Tr_2(\delta_{ij}G_{\sigma_0}\cdot 1\otimes J_j),
\label{solofeom}
\ee
where the subscript on the trace indicates that it operates 
on the second factor in the tensor product and 
\be
G_{\sigma_0}\equiv\frac{1}{kM}\sum_{l=0}^{M-1}\sum_{m=-l}^l
   \frac{\tlm\otimes\tlm^{\dagger}}
        {\frac{l(l+1)}{r^2}+\mu_0^2+g_0\sigma_0}.
\label{fuzzyvectortwopoint}
\ee
Thus using (\ref{solofeom}), we obtain in the large-$N$ limit, 
\be
W[J]=-\frac{k^2}{2}\Tr_{1,2}(\delta_{ij}G_{\sigma_0}\cdot J_i\otimes J_j)
     +W[J=0],
\ee
which leads to the two-point function 
\be
-\frac{\delta W[J]}{k\delta J_ik\delta J_j}=\delta_{ij}G_{\sigma_0}.
\label{fuzzytwopointfun}
\ee
        
Next let us consider the gap equation. Setting $J_i=0$, the partition 
function (\ref{fuzzyvectorpartitionfun}) becomes 
\be
Z=\int\meas{\sigma}\exp(-\frac{N}{2}S_{\ind{eff}}),
\ee
\be
S_{\ind{eff}}=k\Tr\left[-\frac{1}{2}g_0\sigma^2
             +\frac{1}{kM}\log\frac{\det(k(-\D+\mu_0^2+g_0\sigma))}
                                   {\det(k(-\D+\mu_0^2)}\right].
\ee
Let $\sigma_0$ be the above-mentioned scalar matrix, it satisfies 
the saddle point equation 
\be
0=\left.\frac{\delta S_{\ind{eff}}}{\delta\sigma}\right|_{\sigma=\sigma_0}
 =-g_0\sigma_0+\frac{g_0}{kM}\Tr\frac{1}{-\D+\mu_0^2+g_0\sigma_0}\cdot
        \mbox{\bf 1}.
\ee
Thus we have the gap equation 
\be
\sigma_0=\frac{1}{kM}\Tr\frac{1}{-\D+\mu_0^2+g_0\sigma_0}\cdot
        \mbox{\bf 1}.
\label{fuzzygapeq}
\ee
This equation can be regarded as the noncommutative analog of 
the gap equation (\ref{spheregapeq}) of the vector model on the sphere. 
In fact, it is easy to find that (\ref{fuzzygapeq}) leads to 
(\ref{spheregapeq}) in the commutative limit.

\subsection{Master Matrix} 

In this subsection we construct the master matrix which reproduces 
the two-point function calculated above. Applying the master field 
equation (\ref{mastereq}) to the present case, the master matrix 
is a solution to 
\bea
0 & = & -\D\phi_i+\mu_0^2\phi_i+\frac{g_0}{2N}(\phi_i\phi_j^2+\phi_j^2\phi_i)
    +\sum_{\alpha,\beta}M^{-1\alpha\beta}[\phi]
     (\tau^{\alpha}\tau^{\beta}\phi)_i, \label{fuzzymastereq} \\
  & & M^{\alpha\beta}=-k\Tr\left[\phi_i(\tau^{\alpha}\tau^{\beta}\phi)_i
                        \right],~~~~
\tau^{\alpha}:\mbox{infinitesimal generator of $O(N)$}. \nono
\eea
In order to solve this equation, we adopt a following ansatz: \\ 
the master matrix is obtained by replacing $\sph$ with $\tlm$ in the master 
field constructed in section 4. Namely, in an appropriate base, 
the master matrix $\bphi_i$ is given by  

for $l^2 < i < (l+1)^2~(0\leq l\leq M-1)$,
\be
\bphi_i=\left\{\begin{array}{ll}
\sqrt{2}c_l\left.\left(T_{l~\frac{1}{2}\left\{(l+1)^2-i\right\}}
                      +T_{l~\frac{1}{2}\left\{(l+1)^2-i\right\}}^{\dagger}
                 \right)
           \right/2
& \mbox{if $i-l^2:~$odd} \\          
\sqrt{2}c_l\left.\left(T_{l~\frac{1}{2}\left\{(l+1)^2-(i-1)\right\}}
                     -T_{l~\frac{1}{2}\left\{(l+1)^2-(i-1)\right\}}^{\dagger}
                \right)
           \right/(2i)
& \mbox{if $i-l^2:~$even},          
\end{array}
\right.
\label{fuzzymasterdef1}
\ee

for $i=(l+1)^2$,
\be
\bphi_{(l+1)^2}=c_l T_{l0},
\label{fuzzymasterdef2}
\ee

and for $i>M^2$, 
\be
\bphi_i=0,
\ee
where $\{\tlm\}$ is the orthonormal basis defined in section 3.
Note that $T_{l0}$ is a Hermitian matrix.
To be more explicit, for large $N$, 
\be
\bphi=\sqrt{2}\left(\begin{array}{c}
              \left.c_0T_{00}\right/\sqrt{2} \\ \hline 
              \left.c_1(T_{11}+T_{11}^{\dagger})\right/2 \\
              \left.c_1(T_{11}-T_{11}^{\dagger})\right/(2i) \\
              \left.c_1T_{10}\right/\sqrt{2} \\ \hline 
              \left.c_2(T_{22}+T_{22}^{\dagger})\right/2 \\
              \left.c_2(T_{22}-T_{22}^{\dagger})\right/(2i) \\
              \left.c_2(T_{21}+T_{21}^{\dagger})\right/2 \\
              \left.c_2(T_{21}-T_{21}^{\dagger})\right/(2i) \\
              \left.c_2T_{20}\right/\sqrt{2} \\ \hline
                              .          \\
                              .          \\
                              .          \\
              \left.c_{M-1}(T_{M-1\,1}+T_{M-1\,1}^{\dagger})\right/2 \\
              \left.c_{M-1}(T_{M-1\,1}-T_{M-1\,1}^{\dagger})\right/(2i) \\   
              \left.c_{M-1}T_{M-1\,0}\right/\sqrt{2} \\  \hline             
                              0          \\
                              .          \\
                              .          \\
                              .         
                    \end{array}
              \right).
\label{fuzzymaster}
\ee 
It is important that the components of this master matrix for $i>M$ are zero 
as explicitly shown in (\ref{fuzzymaster}).   
Therefore, it makes difference between the vector models 
on the ordinary and the fuzzy sphere in such a way that in the latter case 
we do not encounter the subtlety in taking the large-$N$ limit 
mentioned in the last paragraph in section 4. 

$\bphi_i$ is a `rotational invariant' matrix in the sense that 
for an $SO(3)$ rotation $R$, there exists an orthogonal matrix $O \in O(N)$ 
such that 
\be
U(R)\bphi_iU(R)^{-1}=O_{ij}\bphi_j,
\label{fuzzyclaim}
\ee
where $U(R)$ is an $M$-dimensional representation of the rotation $R$. 
This follows from (\ref{tlmrotation}) by picking up the 
$l^2<i\leq (l+1)^2$ components for each $l$. 
Substituting this ansatz into (\ref{fuzzymastereq}) and considering 
the $l^2<i\leq (l+1)^2$ component, we have 
\be
0 =  -\D\bphi_i+\mu_0^2\bphi_i
        +\frac{g_0}{2N}(\bphi_i\bphi_j^2+\bphi_j^2\bphi_i)
    +\sum_{\alpha,\beta}M^{-1\alpha\beta}[\bphi]
     (\tau^{\alpha}\tau^{\beta}\bphi)_i.
\ee
It is important that, similarly to (\ref{point}), $\bphi_i^2$ satisfies 
the following equation: 
\be
\bphi_i^2=\sum_{l=0}^{M-1}c_l^2(2l+1)\cdot{\bf 1},
\label{fuzzypoint}
\ee
where we used the fact $\tlm^{\dagger}=(-)^mT_{l\,-m}$ and the following 
formula analogous to (\ref{wellknownformula}):
\be
\sum_{m=-l}^l\tlm^{\dagger}\tlm=(2l+1)\cdot{\bf 1}.
\ee
Then the master field equation leads to the following equation 
for $c_l$:
\be
\left(\frac{l(l+1)}{r^2}+2V'(\sum_{k=0}^{M-1}(2k+1)c_k^2)\right)c_l
=\frac{1}{kM}\left(\sum_k\frac{(2k+1)c_l}{c_l^2+c_k^2}
                              -\frac{1}{2c_l}\right),
\ee
where it is used that 
\be
M^{\alpha\beta}=-k\Tr\bphi_i(\tau^{\alpha}\tau^{\beta}\bphi)_i
               =-kM(c_l^2+c_k^2)\delta^{\alpha\beta},
\ee
which can be shown in the same way as (\ref{projection}) and 
$V(x)$ is given by (\ref{potentialform}).
We find in the large-$N$ limit, $c_l$ is given by 
\be
c_l^2=\frac{N}{kM}
      \frac{1}{\frac{l(l+1)}{r^2}+2V'(\sum_l(2l+1)c_l^2)}.
\label{fuzzycoefficient}
\ee
Thus again we obtain the gap equation in the following form: 
\bea
\sigma & = & \frac{1}{kM}
             \sum_{i=1}^{N}\frac{1}
                                 {\frac{l(l+1)}{r^2}+2V'(N\sigma)} \nono \\
       & = & \frac{1}{kM}
             \sum_{l=0}^{M-1}\frac{2l+1}
                                  {\frac{l(l+1)}{r^2}+\mu_0^2+g_o\sigma}.
\eea
where $\sigma=\sum_{l=0}^{M-1}(2l+1)c_l^2/N$. This equation again 
agrees with the gap equation (\ref{fuzzygapeq}). Given the solution 
to the gap equation $\sigma_0$, the master matrix (\ref{fuzzymaster}) 
indeed reproduces the $O(N)$ invariant two-point function  
\be
\frac{1}{N}\sum_i\vev{\phi_i\phi_i}
\equiv -\frac{1}{2N}\sum_i\frac{\delta W[J]}{k\delta J_ik\delta J_i}
=G_{\sigma_0}=\frac{1}{N}\sum_i\bphi_i\otimes\bphi_i,
\label{fuzzytwopointfunction} 
\ee
where we used eqs. (\ref{fuzzytwopointfun}), (\ref{fuzzycoefficient}). 
Moreover, it can be checked that the rules analogous to the Feynman 
diagram expansion hold even in the noncommutative case.\footnote
{For example, performing the perturbative expansion in terms of $g_0$,  
it is found that the `connected' two-point function 
$\vev{\phi_i\phi_j}_{pq,rs}$ includes a following term 
in the order of $g_0^1$, 
$$
-2g_0kM\delta_{ij}\left(\frac{1}{M}\Tr\left(p(G_0)\right)\right)
(G_0\ast G_0)_{pq,rs},
$$
where $G_0$ is a free propagator (\ref{fuzzyfreepropagator}), 
$\ast$ is defined as $(X\ast Y)_{pq,rs}=1/M\sum_{tu}X_{pq,tu}Y_{ut,rs}$ and 
$p$ is a map $p:\mat{M}\otimes\mat{M}\ni f\otimes g \mapsto fg\in\mat{M}$. 
It tends to in the commutative limit, 
$$
-2g_0\delta_{ij}\D_{\ind{F}}(0)\sint\meas{z}\D_{\ind{F}}(x_1-z)
                                            \D_{\ind{F}}(z-x_2).
$$
}      
Therefore, by means of the standard argument based on the Feynman diagram 
expansion, we can show that the theory considered in this section 
also has the factorization property and that all $O(N)$ invariant correlation 
functions can be made from the two-point function 
(\ref{fuzzytwopointfunction}). 
Thus we can conclude that (\ref{fuzzymaster}) is a master field 
of the theory.   

Finally we wish to mention some important points as to the master matrix. 
In view of (\ref{fuzzymaster}), we find that the internal $O(N)$ symmetry 
index $i$ is again associated with the pair of $(l,m)$. 
Although $(l,m)$ is now no more than the label of the matrices  
which form the basis of the representation of $SU(2)$, it can be regarded 
as the `fuzzy space-time' index which tends to the angular momentum 
in the commutative limit. Thus the master matrix mixes the internal 
symmetry and the fuzzy space-time rotational symmetry.
 
As stated in section 3, the fuzzy sphere is regularization 
which manifestly preserves the rotational symmetry.  This is crucial 
for the construction of the master matrix because the master matrix  
(\ref{fuzzymaster}) is composed of the representations of the 
$SO(3)$ rotation. Therefore, it seems that the fuzzy sphere provides  
a sort of regularization scheme suited to formulate the master field.       
\section{Conclusions and Discussions}
\setcounter{equation}{0}

We considered the large-$N$ limit of the $O(N)$ symmetric vector model 
both on ordinary and fuzzy sphere and found that both theories 
have master fields. We also gave the explicit formulas of them 
and observed that they connect  the internal $O(N)$ symmetry 
with (fuzzy) space-time symmetry. We also pointed out that the ultraviolet 
cutoff brought by the fuzzy sphere is essential for the existence 
of the master field in the large-$N$ limit. 

We can think of several possible extensions and applications of 
our work. 

It is interesting to define our model on other fuzzy surfaces and 
observe the relation between the internal symmetry and the fuzzy 
space-time symmetry via the master field. 
Among others, we would like to mention the extension to the fuzzy torus. 
A fuzzy version of the torus \cite{fuzzytorus} is constructed  
by introducing matrices $U,V$ which satisfy the Weyl relation
\be
UV=qVU,
\ee 
as well as the constraints 
\be
U^M=1,~~~~V^M=1,~~~~q=\e^{2\pi i/M}.
\label{torusconstraint}
\ee
Like the fuzzy sphere, the fuzzy torus provides an ultraviolet cutoff 
of the space-time momentum. It is well known that the master field 
of the $O(N)$ symmetric vector model defined in a box with periodic 
boundary conditions exists and links the internal symmetry index 
to the space-time momentum \cite{itoyone,collective}. 
Since Fourier modes in the two-dimensional box correspond to the powers 
of the matrices $U,V$ on the fuzzy sphere, we can expect that 
there exists the master field of the $O(N)$ vector model on the fuzzy torus 
and that it associates the internal symmetry index with the powers of the 
matrices $U,V$. 

Unfortunately, it is only in the particular cases of the sphere and the torus 
that the fuzzy versions of compact surfaces are known. It is important 
to search for a matrix description of the fuzzy surface of higher genus. 

We have considered the model with $M$ fixed and $N\rightarrow\infty$ 
for the purpose of the search for the master fields on the noncommutative 
geometry. However, from the field theoretic point of view, we should 
take the limit $M\rightarrow\infty$ as $M$ plays a role of the cutoff. 
Therefore, it is worthwhile to investigate the model in taking the 
large-$N$ and large-$M$ limit simultaneously in a suitable way 
and to compare it with the double scaling limit in $O(N)$ vector model 
in two dimension \cite{dslofon}.

Recently, `fuzzy supersphere' is proposed in \cite{superfs}. 
It is remarkable that the fuzzy supersphere is found to provide 
the regularization which manifestly preserves supersymmetry. 
It would be interesting to define our models on the fuzzy supersphere 
and to examine the existence of a `master superfield' and the connection 
between the internal and the space-time symmetry, in particular, 
supersymmetry.      

In relation to the nonperturbative formulation of string theory, it is 
shown that from the Matrix theory \cite{BFSS,Matrix} or IIB matrix 
models \cite{IKKT,yoneya,IIBMM} point of view, the noncommutative torus 
appears naturally on the same footing as the standard torus 
\cite{connesetal}. In \cite{douglasetal}, it is also shown that 
gauge theories on noncommutative tori naturally appear as D-brane 
world-volume theories. It would be useful to apply our idea of 
the master field to these theories. 

\section*{Acknowledgements}
I would like to thank Mitsuhiro Kato, Yuji Okawa and Tamiaki Yoneya 
for enlightening discussions.
\newpage

\end{document}